\documentclass[traditabstract]{aa} 
\usepackage{graphicx}
\usepackage{txfonts}
\usepackage{rotating}
\usepackage{natbib}
\bibpunct{(}{)}{;}{a}{}{,}
\renewcommand{\sp}{\hspace{1mm}}
\begin{document}
\title{Giant pulses from the Crab pulsar}

\subtitle{A wide-band study}

\author{R. Karuppusamy\inst{1,3}
  \and
  B. W. Stappers\inst{2,3}
  \and
  W. van Straten\inst{4}
}

\institute{Sterrenkunde Instituut Anton Pannenkoek, 
  University of Amsterdam, Kruislaan 403, Amsterdam, The Netherlands\\
 \email{ramesh.karuppusamy@gmail.com}
 \and Jodrell Bank Centre for Astrophysics, School of Physics and Astronomy, 
  The University of Manchester, Manchester M13 9PL, UK.\\
  \email{Ben.Stappers@manchester.ac.uk}
  \and
  Stichting ASTRON, Postbus 2, 7990 AA, Dwingeloo, The Netherlands
  \and
  Centre for Astrophysics and Supercomputing, Swinburne University of Technology, 
  Hawthorn, VIC 3122, Australia\\
  \email{vanstraten.willem@gmail.com}
}
\date{}

\abstract { The Crab pulsar is well-known for its anomalous giant
  radio pulse emission. Past studies have concentrated only on the
  very bright pulses or were insensitive to the faint end of the giant
  pulse luminosity distribution. With our new instrumentation offering
  a large bandwidth and high time resolution combined with the narrow
  radio beam of the Westerbork Synthesis Radio Telescope (WSRT), we
  seek to probe the weak giant pulse emission regime.  The WSRT was
  used in a phased array mode, resolving a large fraction of the Crab
  nebula. The resulting pulsar signal was recorded using the PuMa II
  pulsar backend and then coherently dedispersed and searched for
  giant pulse emission. After careful flux calibration, the data were
  analysed to study the giant pulse properties. The analysis includes
  the distributions of the measured pulse widths, intensities,
  energies, and scattering times. The weak giant pulses are shown to
  form a separate part of the intensity distribution. The large number
  of giant pulses detected were used to analyse scattering and
  scintillation in giant pulses. We report for the first time the
  detection of giant pulse emission at both the main- and interpulse
  phases within a single rotation period. The rate of detection is
  consistent with the appearance of pulses at either pulse phase as
  being independent.  These pulse pairs were used to examine the
  scintillation timescales within a single pulse period.  }


\maketitle

\section{Introduction}
\label{intro}

Identified as the supernova remnant that resulted from SN 1054, the
Crab nebula is one of the strongest radio sources in the sky, and it
harbours the young neutron star PSR B0531+21. The pulsar is visible
across the entire observable electromagnetic spectrum, and at radio
wavelengths it is the second brightest pulsar in the northern sky. PSR
B0531+21 was discovered by \citet{sr68}, soon after the discovery of
pulsars. This pulsar is noted for several features including the near
orthogonal alignment of the magnetic and rotational axis that gives
rise to the observed interpulse emission. The average emission profile
of the pulsar, obtained by averaging the radio emission from many
rotations of the star, exhibits a number of features that change quite
remarkably with radio frequency \citep{mh94}. The single pulses show a
large variation in amplitude and duration as a function of time. The
most enigmatic of these are its occassional intense bursts known as
giant pulses \citep{hcr70,ss70}. The giant pulses can be extremely
narrow, of the order of 0.4 $n$s \citep{he07} and the pulse flux can
be several 1000 times the average pulse flux. The ultrashort durations
of the giant pulses imply very high equivalent brightness temperatures
\citep{hkwe03} indicating that they originate from nonthermal,
coherent emission processes. In this work, we define giant
  pulses as the pulses with a significantly narrower width than the
  average emission and contain a flux of at least 10 times the mean
  flux density of the pulsar.

The Crab pulsar is one of just a handful of pulsars that have been
shown to have giant pulse emission. Some other pulsars, like the young
Vela pulsar, also show narrow, bursty emission called giant
micropulses \citep{jvkb01}. The fluxes of these micropulses are within
a factor of 3 times the average pulse flux. In the pulsars that show
giant pulse emission, the pulse intensity and energy distributions
exhibit power-law statistics \citep{ag72}, while the giant micropulses
give rise to log-normal distributions \citep{cjd01}. In contrast, the
bulk of the pulsar population have pulse intensities and energies that
follow either a normal or an exponential distribution
\citep{hw74,rit76}. This indicates that the giant pulses and
micropulses may form a different emission population.

The Crab giant pulses have been studied by different groups, yet the
nature of the emission process remains elusive. In the very early
studies at low sky frequencies, the data were afflicted by dispersion
smearing and scattering \citep{hcr70,ga72}, but the power-law nature
of the intensity distribution of giant pulses was identified. In the
next major study, \citet{lcu+95} discuss a multi-wavelength
observation of giant pulse emission, and note the possibility of a
weak giant pulse emission population at radio wavelengths, which they
are unable to resolve owing to insufficient
sensitivity. \citet{sbh+99} found that the Crab giant pulses are broad
band at radio wavelengths. They also determine giant pulse spectral
indices in the range of -2.2 to -4.9 using their widely spaced
observation bands and 29 simultaneously detected giant
pulses. Observations by \citet{hkwe03} revealed that giant pulses at
5.5 GHz contain nanosecond wide subpulses and the presence of such
narrow features has been predicted in numerical modelling by
\citet{wea98}. At these frequencies the radio emission character of
the Crab pulsar changes, with the interpulse emission becoming
dominant. A multi-wavelength radio observation of Crab giant pulses
with widely spaced frequency bands (0.43 GHz and 8.8 GHz) is presented
by \citet{cbh+04}, who discuss the effects of scintillation over a
wide range of frequencies. \citet{ps07} and \citet{eah+02}
investigated pulse width distributions and find that narrow pulses
tend to be brighter. \citet{btk08} carried out a similar analysis in
addition to scattering and dispersion variations in the nebula. All of
these studies point to the peculiarity of the Crab pulsar and its
puzzling emission process, and motivates further study in finer detail
using a large number of pulses. For the work discussed in this paper,
we utilised the wide band capabilities of the new pulsar machine,
PuMa--II \citep{kss08} and the Westerbork Synthesis Radio Telescope
(WSRT) in the coherent tied-array mode.  At small hour angles, the
synthesised beam of the WSRT effectively resolves out the Crab nebula,
reducing the nebular contribution to the system temperature. Thus the
WSRT and PuMa--II combination makes this study much more sensitive in
terms of signal-to-noise ratio achieved, and in number of pulses than
was possible in the past. The rest of the paper is organised as
follows: in \S{2} we describe the observational set up and data
reduction, flux calibration is discussed in \S{3}, the giant pulse
characteristics are discussed in \S{4}.  We report detections of
double giant pulses in \S{5}, and the scattering analysis is presented
in \S\ref{scatter}.

\begin{figure}[htbp]
  \includegraphics[scale=0.95]{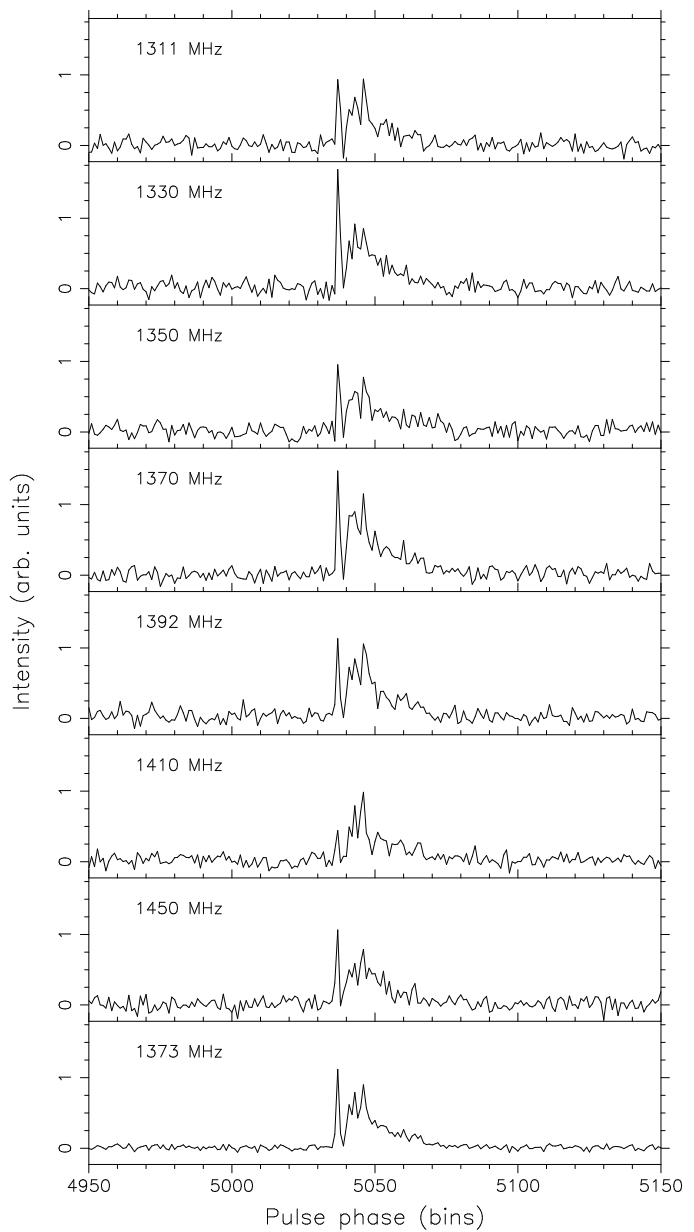}
  \caption{Total intensity of a coherently dedispersed giant pulse at the main
   pulse phase detected in all recorded bands at 4.1 $\mu$s resolution. The
   total dispersion delay of 24.9 $m$s across the seven bands was removed for
   this plot. The lower most panel shows the pulse after combining the signal
   in all seven bands. The pulses displayed here are scaled relative to the pulse at 1330 MHz.}
  \label{GPs}
\end{figure}

\section{Observations and data reduction}
\label{obs}

The radio observations of the Crab pulsar reported here were carried out as 
part of a multi-wavelength observation with the Integral $\gamma$-ray
telescope and the WSRT on 11 October 2005. The WSRT observations were from UTC
$03\fh56\fm50\fs$ to $09\fh36\fm20\fs$ with a break of three minutes in the
middle of the observation to switch data disks. The results of the
$\gamma$-ray observations will be reported elsewhere.

\begin{table}[b]
  
  $$ 
  \begin{array}{p{0.5\linewidth}l}
    \hline
    \noalign{\smallskip}
    Parameter      &  $Value$ \\
    \noalign{\smallskip}
    \hline
    \noalign{\smallskip}
    Observation duration \dotfill & 21420 \sp$s$     \\
    Start Epoch \dotfill       & 53654.726505\sp $(MJD)$\\
    Sky frequencies \dotfill  & 1311^{\mathrm{a}},1330,1350,1370,1392^{\mathrm{a}} \\
    & 1410,1428^{\mathrm{a,b}},1450 \sp $MHz$ \\
    Bandwidth  \dotfill       & 8 \times 20\sp $MHz$ \\
    Nominal $T_{sys}$   \dotfill       & 30\sp $K$ \\
    Beam size  \dotfill      & 21 \arcsec \times 1741 \arcsec\sp ^{\mathrm{c}}\\
    \noalign{\smallskip}
    \hline
  \end{array}
  $$ 
  \begin{list}{}{}
  \item[$^{\mathrm{a}}$] These frequencies are not uniformly spaced to
    avoid interference.
  \item[$^{\mathrm{b}}$] This band was not recorded due to disk failure.
  \item[$^{\mathrm{c}}$] The beam size varies as a function of the observation
    time. See text for details.
  \end{list}
  \caption[]{Telescope parameters and observation details.}
\label{ObsDetails}
\end{table}

\begin{figure}[htbp]
\includegraphics[scale=0.55,angle=-90]{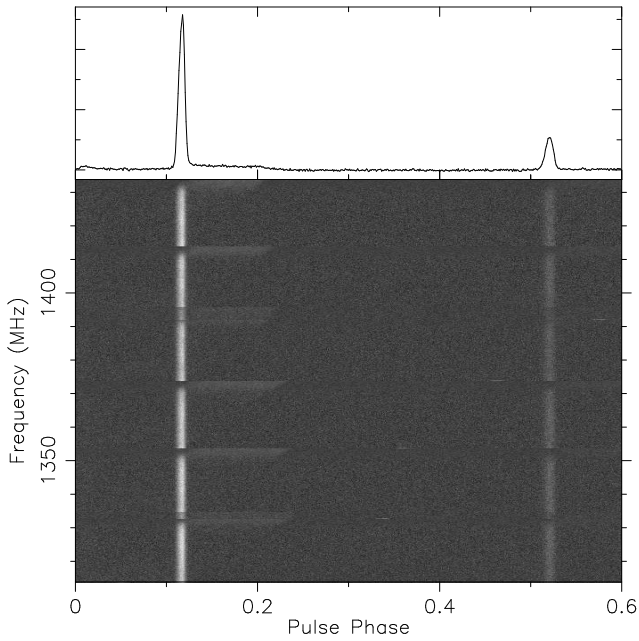}
\caption{ The plot shows the average pulse profile (top panel)
    and the total intensity for six of the seven recorded bands
  in greyscale (lower panel). The striped nature of channels
  at 1330 MHz and 1390 MHz comes from the overlap in the adajcent
  frequency bands. The roll-off of the filters used in the system is
  also seen as a reduced intensity at the band edges. A low-level
  extended feature is seen at the edge (also visible in the top panel
  as the elevated baseline in the right side of the main pulse) of
  each band which is due to the 2-bit quantisation noise and is only
  visible in long exposures.}
\label{FreqPhas}
\end{figure}

The pulsar was observed at eight different sky frequencies in the
L--Band, which is the most sensitive front-end receiver at the WSRT
($T_{sys}=30 $\sp K). The sky frequencies (see
Table. \ref{ObsDetails}) were chosen to be free of radio frequency
interference. Two orthogonal polarisations of 8$\times$20 MHz analogue
signals from each telescope were 2-bit sampled at the Nyquist rate of
40 MHz. The telescope was operated in the tied-array mode in which
coherent sums of the sampled voltages were formed in dedicated adder
units resulting in 6-bit summed voltages. A coherent sum was achieved
by determining the instrumental phase offsets between the telescopes
using observations of a strong calibrator source. These phase offsets,
combined with the geometrical phase offsets required for tracking the
source are applied to each telescope. The resulting values were then
read off as 8-bit data and recorded in the PuMa--II storage
nodes. This resulted in a total of 13.5 Terabytes of raw data. After
the observation, the data were processed offline using the open-source
pulsar data processing software package
DSPSR\footnote{http://dspsr.sourceforge.net/}. A 32-channel synthetic
coherent filterbank was formed across each 20 MHz band with coherent
dedispersion applied across each of the channels using the dispersion
measure (DM) of the pulsar. We obtained the DM ($=56.742$) from the
Crab pulsar ephemeris maintained by the Jodrell Bank
Observatory\footnote{http://www.jb.man.ac.uk/~pulsar/crab.html}
\citep{lps93} at the epoch closest to our observation. Frequency
resolution was preserved so that studies of spectral indices,
scintillation, and scattering could be carried out.

The total intensity was computed for each pulse from the dedispersed
data. Giant pulses were detected by computing the peak signal-to-noise
ratio (denoted by $S/N$). The giant pulse detection threshold was set
at $S/N\ge7\sigma$ in each band, where $\sigma$ is the off-pulse
root-mean-square noise fluctuation. Pulses below the detection
threshold were discarded to ease storage requirements. The original
sampling time was 25 $n$s. The 32-channel filterbank and the choice of
4.1 $\mu$s final time resolution resulted in 8192 phase bins. The time
resolution of 4.1 $\mu$s was chosen to match the estimated scattering
timescale available at the time \citep{sbh+99}. However, it is known
from recent work by \citet{btk08} that single pulses at these radio
frequencies can be as narrow as 0.5 $\mu$s. In addition to the single
pulses, average pulse profiles with 128 frequency channels in each
20\,MHz band were formed every 10 seconds.

The reduced data consisted of $\sim$21000 giant pulse candidates in
each recorded band. An example candidate is shown in Fig. \ref{GPs},
where the pulse was detected in all bands. In the offline analysis
stage, these candidates were combined in software using only pulses
that show the expected dispersion delay. This method ensures that
spurious signals were filtered out in our analysis. After combining in
software, 12959 giant pulses were identifed to have occurred
simultaneously at all observed sky frequencies. Of the 12959 pulses,
11384 were detected at the main pulse phase and 1370 at the interpulse
phase of the average pulse profile. 

The data were folded and the single pulses were formed using the DSPSR
software package and a polynomial determined by using TEMPO
\citep{tw89}. The folded profiles formed in each 20-MHz band were
combined in software to validate the DM used. The combined data are
shown in Fig. \ref{FreqPhas} as a frequency--phase image and shows
no smearing, confirming that the value of DM is correct. A similar
procedure was used to combine simultaneous giant pulses in all seven
bands. Some artifacts of the 2-bit systems of the individual
telescopes are visible once the profile is summed for the entire
six-hour long observation. The width of these artifacts match the
dispersion smearing in the bands as seen in the top panel of Fig.
\ref{FreqPhas}. The quantisation noise is 12\% for a single telescope,
whose signal is sampled using 2-bits \citep{coo70}. Since signals from
the 14 telescopes of the array were coherently summed, the
uncorrelated quantisation noise was reduced by a factor of
$\sqrt{14}$. The resulting noise of 3.7\% is considered too small to
be problematic in the analysis that follows. In many stages of the
analysis, extensive use of the PSRCHIVE \citep{hvm04} utilities was
made to view and validate the pulsar data and to compute the $S/N$ used
in later analysis.

\section{Flux calibration}
\label{fcal}
To establish a flux scale for the observed giant pulses, the mean system flux
needs to be computed. The mean system flux is proportional to the r.m.s noise
variations at the telescope output and can be expressed by the radiometer
equation \citep{dic46},

\begin{equation}
  S_{min} = \frac {S_{sys}}{\sqrt{N_{p}\cdot B \cdot T_{int}}}\,,
\label{sav} 
\end{equation} 

\noindent where, $S_{min}$ is the r.m.s system noise in Jy, $S_{sys}$
the total system noise, $N_p$ the number of polarisations ($=2$), $B$
the bandwidth in MHz ($=140$), and $T_{int}$ the integration time in
seconds.
\begin{figure}[htbp]
  \includegraphics[scale=0.65,angle=-90]{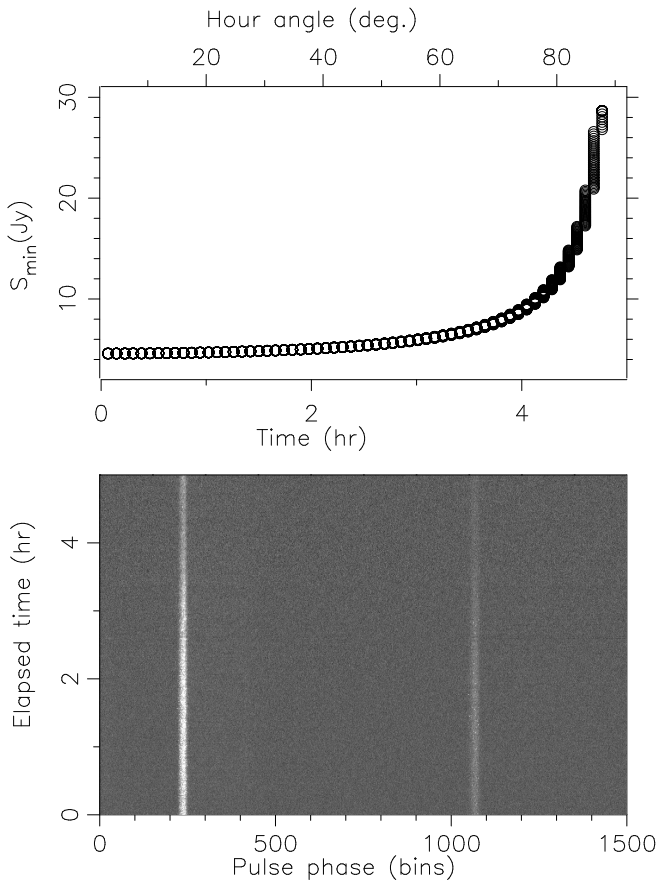}
  \caption{The upper panel shows the change in minimum detectable
  signal $S_{min}$ within a 4.1 $\mu$s time interval during the first
  5 hours of the 6-hour observation. The hour angle of the source is
  displayed on the top ordinate axis. The lower panel is the observed
  change in peak signal-to-noise ratio of the detected signal. The
  dependence of the signal-to-noise ratio on the hour angle of source
  is discussed in the text.}
  \label{Fsys}
\end{figure}
The total system noise in flux density units in eqn. (\ref {sav}) is the total
equivalent system temperature divided by the telescope gain
($S_{sys}=T_{total}/G$). For the WSRT, when signals from the fourteen 25-m
parabolic dishes are combined, the resulting telescope gain\footnote{The
telescope gain is 1.34 K.Jy$^{-1}$ for an ideal array combiner. The reduction
in gain is attributed to losses in the formation of the tied-array signal.} is
$G=1.2$ K.Jy$^{-1}$. The term $T_{total}$, can again be expressed as
\begin{equation}
  T_{total} = T_{sys} + f_{\nu}(t)\cdot T_{cn}\,.
\label{t_tot} 
\end{equation}
The term $T_{cn}$ is the contribution of the Crab nebula to the system
temperature, while $f_{\nu}(t)$ is a time-dependent factor explained
below. Following \citet{bkf+97}, we express the total flux of the Crab
nebula at frequency $\nu$ (in GHz) as $S_{CN}=955\nu^{-0.27}$ Jy, from
which $T_{cn}$ is computed. The WSRT is an east-west array and the
coherent addition of the telescope signals results in a $21 \arcsec
\times 1741 \arcsec$ fan beam. The Crab nebula is an extended source
of size $\Omega_{cn}=6 \arcmin \times 4 \arcmin$, so the WSRT's fan
beam resolves the Crab nebula in the east-west direction. This in turn
reduces the nebular contribution to the T$_{sys}$. However, the width
of the WSRT's fan beam is not a constant, but is a function of the
observation time. While the source is being tracked, the
effective width of the synthesised beam changes with hour
  angle (HA) and it is expressed as $\Omega_A(t)=\Omega_{cn}
\cdot\lambda/D \cdot cos(HA)$. In this expression $HA=t - RA$, where
$t$ is the local sideral time, the maximum baseline $D=2700m$, and
$RA$ is the right ascension of the Crab pulsar. The fraction of the
nebular contribution can be expressed as
$f_{\nu}(t)=\Omega_A(t)/\Omega_{cn}$, which reaches its minimum value
of 0.13 at zenith. As the source is tracked towards the horizon, the
projected distance between the dishes decreases and $\Omega_A(t)$
increases. Consequently, the observing system becomes less sensitive
toward larger hour angles, or when the source rises and sets. This
time dependence of the system noise is included in our flux
calibration. The variation in $S_{min}$ is shown for a bandwidth of
140 MHz, $N_p=2$ and $\tau=4.1\sp\mu$s in the upper panel of Fig.
\ref{Fsys}. A plot of the pulse intensity during the observation
(lower panel of Fig. \ref{Fsys}) confirms this reduction in
sensitivity.

The peak flux of the giant pulses were computed using the modified
radiometer equation \citep{lk05} for the pulsar case,
$S_{peak}=(S/N)\cdot S_{min}$. With the above considerations of the
nebular contribution to $T_{total}$ and with $T_{sys}=30$K in the
WSRT's L--Band, the system retained sufficiently high sensitivity in
the first 15000 seconds of the observation. Two other factors have
been neglected in this calibration procedure and do not contribute
significantly to the $T_{sys}$ : the relative change in the
orientation of the WSRT's fan beam and the Crab nebula over the course
of observation and the partial shadowing of three telescopes out of
the 14 for HA $> 54\degr$ (the last 3 hours of our observation).

\section{Single-pulse statistics}
\label{inten}

For the analysis that follows, all pulses that were flux-calibrated as
described in the previous section were used. The discussed change in
system sensitivity does not limit this analysis thanks to our careful
flux calibration procedure. While approximately 70\% of the pulses
were detected in all seven bands simultaneously, the rest were
detected in two or more of the seven bands recorded. For the results
described below, where applicable, only those pulses that were
detected in all seven bands were used and explicitly mentioned.

\subsection {Pulse intensity distributions}
\label{pintensity}

\begin{figure}[htbp]
  \includegraphics[scale=0.65,angle=-90]{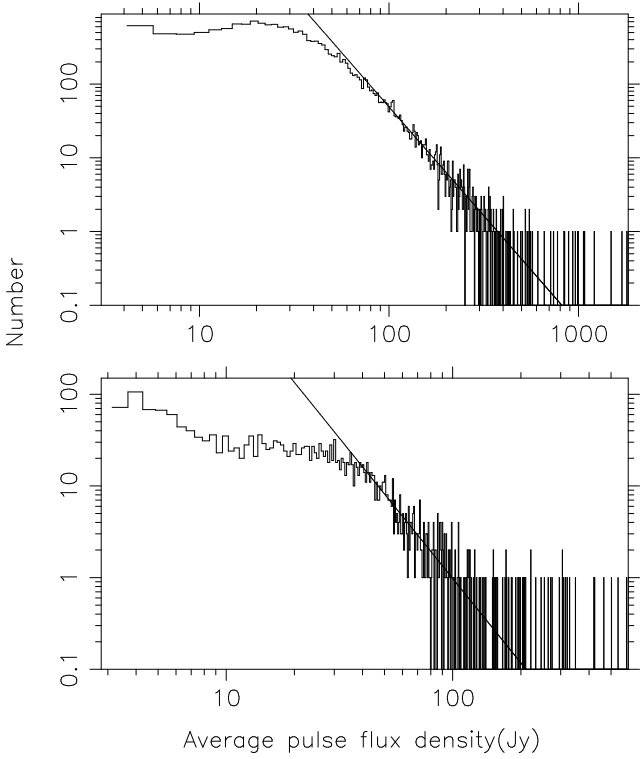}
  \caption{ Distribution of the pulse intensity of all giant pulses
    detected at the main- and interpulse phases in the upper and lower
    panels, respectively. The long tail results from the giant pulse
    emission. The best fit power-law curve is shown with slope
    $-2.79\pm0.01$ for the pulses in main pulse phase and
    $-3.06\pm0.06$ for the pulses in the interpulse phase. Both
    distributions show an excess near 4Jy and come from the rounding
    off in $W_{eq}$. [{\em{see text for details}}].}
  \label{fig2}
\end{figure}

The giant pulse fluxes of the Crab pulsar contribute to the long
exponential tail of the single pulse intensity histograms
\citep{ag72}, while the normal pulsars show Gaussian or exponential
pulse intensity distributions \citep{hw74}. Fig. \ref{fig2} shows the
average pulse flux distribution for pulses detected in at least two of
the seven recorded bands. The average pulse flux is computed by
integrating all emission within the equivalent width, $W_{eq}$ of the
giant pulse (see \S\ref{width}). This value is averaged over the pulse
period to obtain the average pulse flux. The pulse in each band was
detected based on a threshold of $7\sigma$. A pulse detected in two
bands satisfies the $\sqrt{2} \times 7 = 9.89\sigma$ limit. In the
first three hours of the observation (when the system was most
sensitive), the flux equivalent system noise in 4.1 $\mu$s is 109
Jy. Averaged over the pulse period, a pulse of $S/N = 9.89\sigma$
corresponds to an average pulse flux density of 3.9 Jy. This implies
that it is sensitive to all pulses greater than $27\times \langle F
\rangle$, where $\langle F \rangle=14 m$Jy is the average flux density
of the Crab pulsar. Therefore, the flux distribution computed here
contains a good fraction of weak giant pulses compared to those
reported elsewhere (see Table. \ref{flux-tbl}).

The intensity distributions displayed in Fig. \ref{fig2} shows at
least two components: a peak at or below $\sim 4$ Jy -- the weak
pulses that may comprise the trailing part of the normal pulse
distribution. The next component peaking at $\sim 20$ Jy resembles a
lognormal distribution with a power-law tail. The bright giant pulses
result in the extended power-law tail and is described by $ N_{F}
\propto F^{\alpha}$, where $N_{F}$ is the number of pulses detected in
1.8 Jy flux intervals of $F$. The value of $\alpha=-2.79\pm0.01$ and
$\alpha=-3.06\pm0.06$ was determined from the best fits to the data in
the interval 118 Jy $\le F \le$ 2000 Jy and 40 Jy $\le F \le$ 596 Jy
for the giant pulses in the main- and interpulse, respectively. Visual
inspection of Fig. \ref{fig2} shows that the distribution is
multi-modal, with giant pulses in the region $F \ga 10$ Jy and the
pulses below this limit possibly representing normal pulses.

\begin{table}[h]
  $$ 
  \begin{array}{p{0.5\linewidth}cc}
    \hline
    \noalign{\smallskip}
    Reference & $Frequency$  &  $Threshold$ \\
    & $(MHz)$& $(Jy)$\\
    \noalign{\smallskip}
    \hline
    \noalign{\smallskip}
    \citet{lcu+95} \dotfill & 800  & 120.0\sp    \\
    \citet{ps07}  \dotfill  & 1197      & 5.9\sp^{\mathrm{a}}\\
    \citet{btk08} \dotfill  & 1300/1470  & 22.3\sp^{\mathrm{b}} \\
    This paper \dotfill & 1373 & 3.9\sp  \\
    \noalign{\smallskip}
    \hline
  \end{array}
  $$ 
  \begin{list}{}{}
  \item[$^{\mathrm{a}}$] Equivalent average pulse computed flux from
  the quoted 6$\sigma$ peak flux density of 142 Jy, assuming 0.036
  pulse duty cycle.
  \item[$^{\mathrm{b}}$] Average pulse flux density extrapolated for
  $7\sigma$ threshold, 4.1$\mu$s time resolution and pulse duty
  cycle $\approx$ 0.036.
  \end{list}
  \caption[]{Reported sensitivity to the Crab giant pulse observations
  in the literature.}
\label{flux-tbl}
\end{table}

It is worth noting the differences in the intensity distributions
displayed in Fig. \ref{fig2}. While the distribution of the giant
pulses in the main pulse phase shows a clear turn over at $\sim$20 Jy,
the emergence of a bimodality in the region containing weak pulses is
evident in the intensity distribution of the interpulse giants. The
distribution corresponding to the interpulse phase also shows a
flattening in the 10--30 Jy region. The clear excess of weak pulses in
both the distributions in the region $ F\le$ 4Jy is due to our method
of setting $W_{eq}=4.1\sp\mu$s (equal to the time resolution). In this
case the emission window we considered is dominated by noise or weak
and narrow pulses. The slopes of the power-law models obtained here
can be compared to the values reported earlier. Fig. 4 of
\citet{lcu+95} shows a slope of $-3.46\pm0.04$ for data at 800 MHz,
which is slightly steeper than the slopes of the main- and interpulse
distributions derived here. \citet{cbh+04} derive a value of $\sim$
-2.3 at 433 MHz and \citet{btk08} found $-2.33\pm0.14$ at 1300 MHz,
which are comparable to the slope the main pulse intensity
distribution in our work. The slopes of the intensity distribution
reported here generally agree considering the effect of low number
statisics and/or dispersion smearing in the observations reported
elsewhere. While this experiment was sensitive to much lower fluxes,
the long observation time has also enabled the detection of rarer
bright pulses.

\subsection {Pulse energy distributions}
\label{penergy}

\begin{figure}[htbp]
  \includegraphics[scale=0.65,angle=-90]{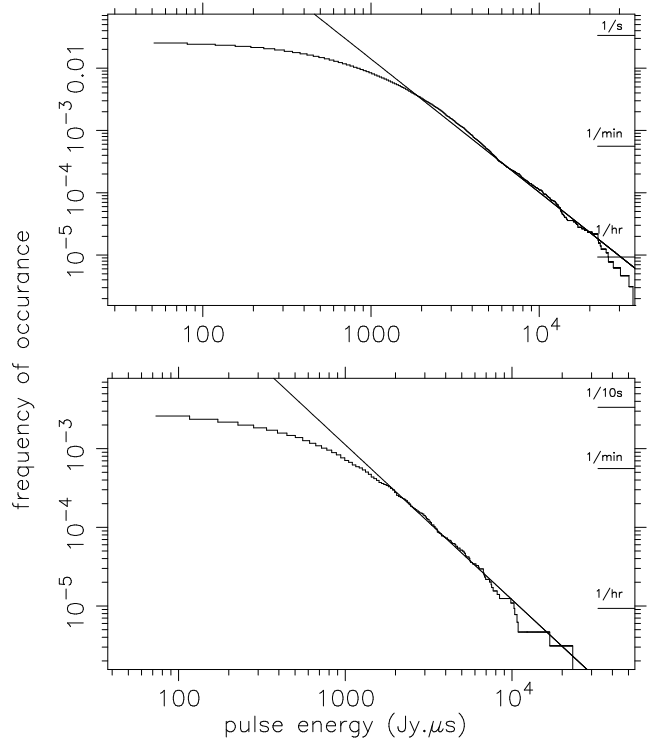}
  \caption{ The cumulative probablity distribution of the energy in
    giant pulses detected at the main pulse and the interpulse phases
    in the upper and lower panels, respectively. The y-axis is the
    fraction of the total number of pulses and pulse energy is plotted
    on the x-axis. Also shown are the occurrence rates per minute,
    second and hour.}
  \label{fig4}
\end{figure}

The relative occurrence rates of giant pulses is displayed as a
cumulative probablity distribution of the individual pulse energies in
Fig. \ref{fig4}.  The pulse energy is computed by multiplying the
equivalent width, $W_{eq}$, and the average pulse flux. As described
in \S{\ref{pintensity}}, we computed the best fits to the cumulative
probablity distributions of the main- and interpulse giants. The
power-law curve with $\alpha=-2.13\pm 0.007$ and $\alpha = -1.97 \pm
0.006$ fits the data for pulse energies at the main- and inter pulse
phases, respectively. The break seen at $\sim$2000 Jy.$\mu$s is
consistent with the break value reported by \citet{ps07}. The emission
at the interpulse phase shows a somewhat shallower power-law.

It is known from \citet{ps07} that the power-law index has a width
dependence, varying from $-$1.7 to $-$3.2 as the pulse width
increases. Based on this variation, the index we find is in good
agreement with \citet{ps07} and \citet{btk08} ($-1.88\pm 0.02$ at 1300
MHz). However, we fit only a single power law unlike the two power-law
fits found by these authors. Partial fits to the low-energy pulses
yield more than two components, with shallower power-law indices
indicating a simple dual-component fit is insufficient. One
explanation for this can be the bias introduced by setting
$W_{eq}=4.1\sp\mu$s for narrow pulses, overestimating the pulse
energy. However, this can only be a minor contribution and is an
argument that there is a clear break in the intensity distribution.
To compare the occurrence rates we see here, we proceed to derive the
rates from the arrival times of the giant pulses in the next section.

\subsection {Giant pulse rates}
\label{rate}

\begin{figure}[htbp]
  \includegraphics[scale=0.65,angle=-90]{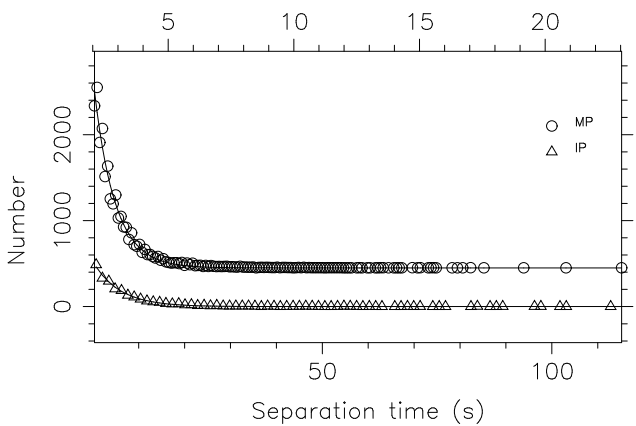}
  \caption{ The symbols show the distribution of separation times between
  successive giant pulses at the main- and interpulse phases and the solid
  lines are the best fits to the distribution. The top ordinate axis
  corresponds to the curve and data for the pulses at the main pulse phase and
  are offset by 450 for clarity.}
  \label{hist_rate}
\end{figure}

The distribution of the separation times between successive giant
pulses is plotted in Fig. \ref{hist_rate}. If the giant pulses are
mutually exclusive events independent of each other, then the arrival
time separation follows a Poisson process \citep{lcu+95}. The
probablity of a giant pulse occurring in the interval $x$ is then
given by $P(x) = \mu x. e^{-\mu x}$, where $\mu$ is the mean pulse
rate. Since our data only consist of giant pulses, we expected to see
an exponential reduction in the separation time between the
pulses. Fig. \ref{hist_rate} shows the fits to the separation times
at both the inter- and main-pulse phases.

Functions with an exponential decay with time constants $1/\tau= 1.1
\pm 0.02$ and $1/\tau= 0.172 \pm 0.003$ are in excellent agreement
with the data at the main- and interpulse phases, respectively. From
the values of $\tau$, the mean giant pulse rates are one main- pulse
giant every 0.9 seconds and one inter pulse giant every 5.81 seconds
observed above our threshold limit of 3.9 Jy. At these frequencies,
the interpulse giants are comparatively less numerous as is evident
from our data. For comparision, the inter-pulse giants are brighter
and more frequent in frequency bands above 5.5 GHz \citep{cbh+04}. The
combined rate of the giant pulses (fit and data not shown) is one
pulse every 0.803 seconds. The foregoing discussion confirms earlier
predictions that the giant pulse rate increases with frequency for the
Crab pulsar \citep{lcu+95,sbh+99}. The effect of the WSRT's
sensitivity reduction towards the end of the observation, as displayed
in Fig. \ref{Fsys}, may have contributed to the long tail of the
distribution, where fewer pulses were detected than in the first half
of the observation. However, the rate derived here is robust, since
the system had sufficiently high sensitivity in the first half of the
observation.

\subsection {Width distributions}
\label{width}

\begin{figure}
\centering
  \includegraphics[scale=0.65,angle=-90]{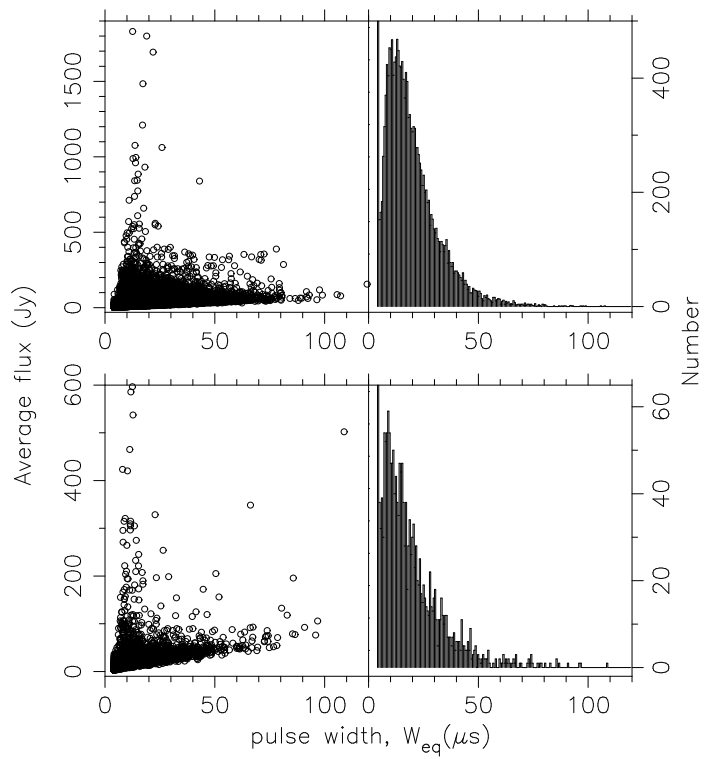}
  \caption{Plot of intensity against pulse width for the main- and
    interpulse windows in the top left and lower left
    panels. Histograms of equivalent pulse widths are shown in the top
    right and lower right panels. The distribution has an exponential
    envelope. For pulses with computed $W{eq}<4.1\sp\mu$s due to
    random noise fluctuations the widths were rounded off to 4.1
    $\mu$s.}
  \label{fig5}
\end{figure}

The equivalent pulse width, $W_{eq}$ is defined as the width of a
top-hat pulse with height equal to the peak intensity of the
pulse. $W_{eq}$ for the giant pulses detected in all seven bands was
computed. The results are displayed in panels on the right in
Fig. \ref{fig5}. We express $W_{eq}$ as

\begin{equation}
\label{Weq}
W_{eq} = \frac{1}{I_{max}} \times \displaystyle\sum_{i=n_1}^{n_2}{I_i} \times 4.1\sp\mu s \,,
\end{equation}

\noindent where $I_{max}$ is the peak intensity, $I_i$ the intensity
in the pulse emission window defined by bins $i=n_1\cdot\cdot\sp n_2$ and is
equal to 1ms in our case. Thus $W_{eq}$ can be viewed as the
equivalent width of a rectangular pulse in $\mu$s that has the same
area as the giant pulse, with height $I_{max}$.

The giant pulses at these frequencies can be quite narrow. For
instance, \citet{btk08} find pulse widths to be 0.5 $\mu$s and
\citet{eah+02} found 0.2 $\mu$s. Our method of data reduction allowed
a time resolution of 4.1 $\mu$s, so pulses with $W_{eq} < 4.1 \mu$s
were taken to have a width equal to 4.1 $\mu$s. This results in some
pulses being underestimated in flux and overestimated in equivalent
width. The computed equivalent widths range from 4.1 $\mu$s to
$\sim$120 $\sp\mu s$, and we find that bright pulses tend to be narrow
as seen in the left hand panels of Fig. \ref{fig5}. This was also
suggested by \citet{sbh+99} and shown by \citet{eah+02}. \citet{ps07}
found a similar behaviour in addition to a width-dependent break in
the power-law fits to the pulse-energy distribution.

In the seven closely spaced radio bands observed, we note that a vast
majority of the pulses have widths larger than 4.1 $\mu$s. This is
seen in the pulse width histograms at the two pulse phases, displayed
in the panels on the right in Fig. \ref{fig5}. The distribution shows
a peak at $\sim$16$\sp\mu$s, which is 4 times our ultimate time
resolution in the main pulse, and the peak shifts towards narrower
timescales for the interpulses. We find less than $9\%$ of the pulses
with $W_{eq}=4.1\sp\mu$s, indicating that the majority of the pulses
show wider widths than our time resolution. The shape of the width
distribution is similar at both the main- and interpulse phases.The
contribution to the tail region of the distribution comes from scatter
broadened pulses.

\subsection{Spectral index of giant pulses}
\label{sindex}

\begin{figure}[hbtp]
  \includegraphics[scale=0.65,angle=-90]{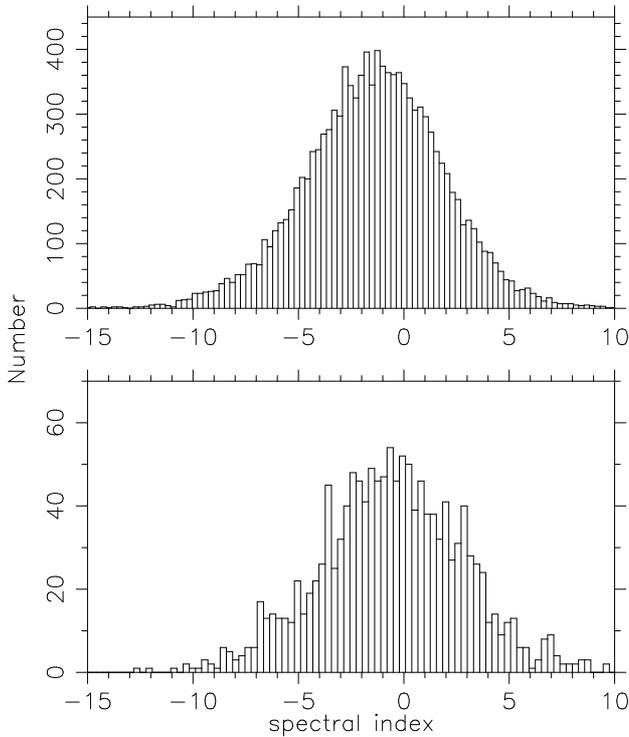}
  \caption{ Histogram of spectral indices for the giant pulses detected at the
  main pulse (bottom panel) and the interpulse phase (top panel). The spread
  in the distributions is indicative of fitting errors. See text for details.}
  \label{IndexHist}
\end{figure}

The data were recorded in 7 different radio bands each 20 MHz wide in
the frequency range 1300--1450 MHz, and several thousands of pulses
were detected simultaneously in all bands. The spectral index of
individual pulses was computed by modelling the flux variation of a
giant pulse as $S(\nu) \propto \nu^k$. Here, $S(\nu)$ is the flux of
the giant pulse at frequency $\nu$, and $k$  the spectral index. The
histograms of the derived spectral indices are displayed in Fig.
\ref{IndexHist} for the giants at both pulse phases. A large
dispersion in the spectral index is seen, with values $-1.44 \pm 3.3$
for the main- and $-0.6 \pm 3.5$ for the interpulse giants.

These spectral index values are quite a bit shallower than those
detected previously (see Introduction) over wider frequency
separations.  We therefore consider the effects of diffractive
interstellar scintillation (DISS) on the spectral index
estimates. Strong DISS results in pulse intensity variations within
each of the seven bands. The effect of scintillation is to modulate
the observed pulsar signal in both time and frequency. This is seen as
regions of enhanced or diminished brightness in a grey scale plot of
the intensity as a function of time and frequency. These regions are
known as scintles. We estimate the scintillation bandwidth based on
the pulse scatter timescales, $\tau_s=395 \pm 50 \mu$s at sky
frequency of 200 MHz, as reported in the work of \citet{bwk+07}.  We
further make use of their revised $\tau_s \propto \nu^{-3.5}$
frequency scaling and consider that the scintillation bandwidth and
scattering timescale are related by $2\pi\Delta\nu_d\tau_s = C_1$ ,
where the constant $C_1=1.05$ for a thin scattering screen
\citep{cbh+04}.  From these considerations $\Delta\nu_d \approx $
0.25--0.38 MHz in the 1300--1460 MHz band. On examining a few giant
pulses by eye, it was clear that some of the scintles are resolved,
while some were narrower than our channel width of $\Delta\nu = 0.625$
MHz.  Thus, in the flux obtained by integrating the signal in the
20MHz-wide bands, the scintles tend to average out. This implies that
scintillation does not cause the spread in the individual
  giant-pulse spectral indices.  Moreover, with such narrow
scintillation bandwidths, averaging over many giant pulse spectral
index determinations as we have done here would give an average
spectral index that reflects the true average spectral index.

Refractive interstellar scintillation (RISS) cannot corrugate the
spectra of single pulses, since the pulse intensity variations due to
RISS are noticeable in observation of the order of a few days
\citep{lcu+95}. However, the pulses do have a significant structure
that is intrinsic to the emission process. One example is displayed in
Fig. \ref{GPs} and these pulses do contribute to the spread in the
computed spectral indices. In this figure, it is clear that the
leading short burst shows considerable variation across the seven
bands, while the scattered trailing part of the pulse is correlated
across frequency. This is again similar to what \citet{he07} find, as
shown in their Fig. 4, but at a much higher frequency of $\sim$9
GHz.

\citet{sbh+99} find that the spectral index variation is between
$-$4.9 and $-$2.2 based on 29 pulses they observed in two bands
centred at 1.4 GHz and 0.6 GHz. The spread in the indices computed
here and that of \citet{sbh+99} points to the stochastic nature of the
giant pulse emission process and/or the disturbed plasma flow in the
magnetosphere caused by strong plasma turbulence \citep{he07}.  The
giant pulses used in this analysis were detected in all seven bands
and represent 70$\%$ of all detected pulses in our data. Since each of
our bands is 20 MHz wide, detection in seven bands implies an emission
bandwidth of at least $\Delta \nu=$ 140 MHz. This suggests that the
emission bandwidth of Crab giant pulses is potentially greater than
$\Delta\nu/\nu = 0.1$, unlike the giant pulse emission from the
millisecond pulsar B1937+21 \citep{ps03}. We note that the
$\Delta\nu/\nu = 0.8$ for the Crab giant pulses reported by
\citet{sbh+99} was based on 29 simultaneous giant pulses from their
90-minute observation ($\sim$ 161086 stellar rotations). Those 29
pulses could have been chance detections, while the $\Delta\nu/\nu =
0.1$ limit derived here comes from a much larger sample of giant
pulses so is more robust. We detected a total of 17587 giant
pulses, of which approximately 4000 were detected in less than 7
bands. Clearly it is impossible to include the pulses detected in only
a few bands in this analysis as that would increase the dispersion in
the spectral indices computed; however, this lack of detection in all
bands, for pulses which were clearly detected in the other bands, is
an argument for there being some narrow band effects that appear to
modulate the giant pulse intensity.

\section {Double giant pulses}
\label{doublegiants}

During direct inspection of some giant pulses, it was noticed that
occasional giant pulse emission was evident at both the main- and
interpulse phases within a single rotation period of the star. To
determine how many such pulses were present, the following search
algorithm was used. First, the giant pulses detected in all seven
bands were combined in software across the frequency bands. The pulses
were then averaged over polarisation and frequency to create single
pulse total intensity profiles. The search algorithm was made
sensitive to emission at both emission windows (main- and interpulse)
by traversing each pulse profile twice; in the first pass, the
emission peak and phase information was recorded, following which a
search is made in the other emission window i.e. if a pulse was
detected at the main pulse phase we check whether a pulse is also seen
at the interpulse phase. All pulses that show signal $\ge 5\sigma$ in
the second emission window are collected separately. The pulses
returned by the search procedure were examined by eye to validate the
double pulse nature.  To our knowledge, this is the first instance of
this phenomena being reported. A total of 197 pulses that show
emission at both pulse phases were found in our data set above the
$5\sigma$ detection threshold.

\begin{figure}[htbp]
  \includegraphics[scale=0.65,angle=-90]{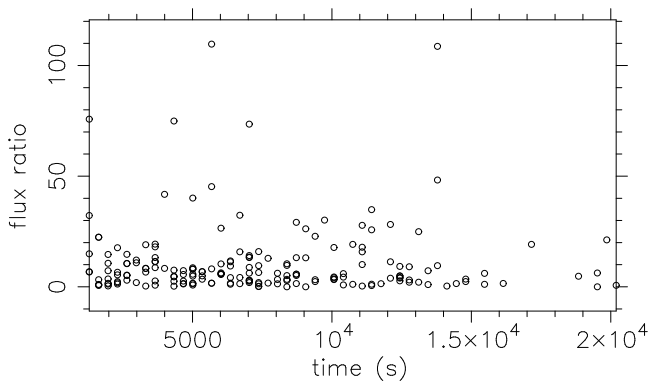}
  \caption{ Detected double giant pulses shown as a ratio of the main
    pulse to the interpulse flux. The x-axis shows time since the
    start of the observation.}
  \label{double_ratio}
\end{figure}

To consider how likely this is to happen by chance, we note that the
observation lasted 643263 rotations of the star and 11584 and 1375
giant pulses were found at the main- and interpulse phases,
respectively, above the $7\sigma$ detection threshold in each
band. Since these giant pulses were detected in all seven bands, the
effective threshold is now $\sqrt{7}\times 7\sigma=18 \sigma$. If the
$18 \sigma$ criterion is used to search for the double pulses, a total
of 17 pulses are seen. In other words, only 17 pulses in the 197
detected show $S/N \ge 18\sigma$ in either of the two emission
windows. Let the giant pulses occurring at the two pulse phases be
independent events, with individual probablitites $P(A)$ and
$P(B)$. The chance of two giant pulses occurring within a single
rotation period is the joint probablity $P(A,B)=P(A).P(B)$. Thus the
chance of detecting a giant pulse above the $18\sigma$ threshold limit
at the main- and interpulse phases are $P(A)=11584/643263$ and $P(B)=
1375/642263$ leading to $P(A,B)= 3.5\times 10^{-5}$. We therefore
expect a total of $P(A,B)\times 643263$ = 24 pulse periods with pulses
at both phases in our data. The detection of 17 pulses is thus
consistent with the expected 24 pulses.

As seen above, combining the seven bands improves sensitivity and
allows the detection of weaker pulses. Considering pulses with $S/N$
greater than $5\sigma$ in the second emission window resulted in the
detection of an additional 180 double pulses. While the 197 pulses
detected are not sufficient to perform meaningful statistics of these
pulses, in \S\sp\ref{subscatter} we use our population of double giant
pulses to study scintillation and scattering within a 0.5 rotation of
the pulsar.

Although the appearance of the pulses in the same rotation period is
consistent with the individual occurrence rates, we compared the GP
properties at each phase. In the double pulses, the emission in the
interpulse phase is typically narrower ($W_{eq}\la 16\sp\mu$s) than
the emission at the main pulse phase and pulses at the main pulse
phase are typically brighter, as shown in Fig. \ref{double_ratio}. In
both cases this is consistent with the known population of GPs at each
phase. A similar analysis to the one in \S\ref{rate} was done to
determine the rate of double pulses and a rate of 1 pulse in 84
seconds, or one in 2545 rotations of the star was found to have giant
pulse emission at both pulse phases. Thus, given the narrowness and
very low occurrence rates of these pulses, they were easily missed in
earlier observations.

\section{Single-pulse scattering}
\label{scatter}

The frequency resolution and large bandwidth of our data benefits
scattering and scintillation checks on the individual pulses in two
ways. First, the pulses detected in 7 bands are combined in software
to give 224 channels across the 140 MHz bandwidth allowing examination
of scintillation. Second, the large bandwidth of the combined pulse
increases sensitivity and makes it possible to identify low-level
extended scatter tails. To characterise the scattering time $\tau_s$
in the pulsar signal, we computed the extent of pulse broadening in
the individual giant pulses. If the pulses are scattered by a
thin-screen between the source and the observer, the pulses can then
be modelled as an one-sided exponential with a vertical rise and a
rapid decay \citep{wil72}. This can be written as

\begin{equation}
\label{scatter-eqn}
f(t) = \left\{ 
\begin{array}{l l}
  e^{-t/\tau_s} & \qquad \mbox{    if   $t \ge 0$ }\\
  0 & \qquad \mbox{    if $t < 0$}.\\ 
\end{array} \right. 
\end{equation}

\begin{figure}[htbp]
  \includegraphics[scale=0.65,angle=-90]{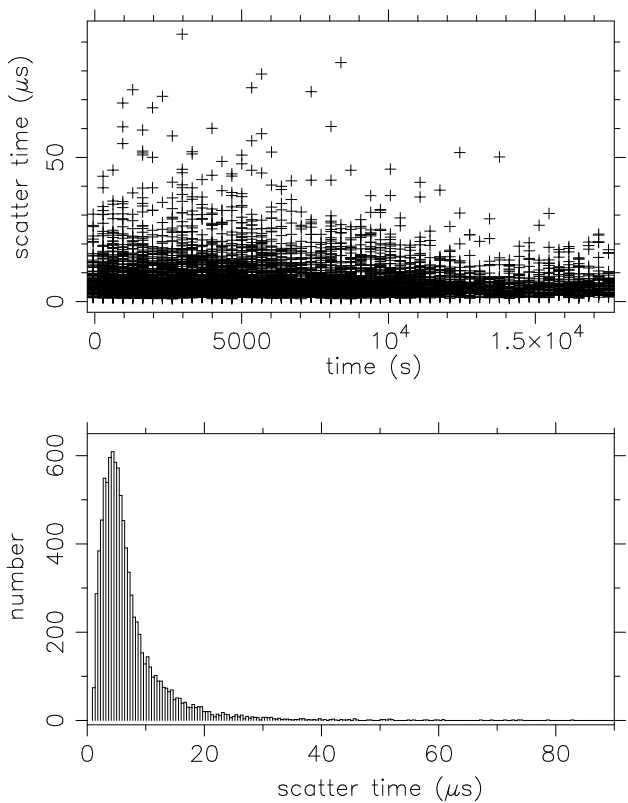}
  \caption{ Upper panel: a plot of the values of time constant $\tau_s
    $ from the fits to the scattering in the entire duration of
    observation for those pulses detected at the main pulse phase. The
    lower panel displays a histogram of the time constants obtained
    from exponential fits to the scatter tails of individual
    pulses. The values of $\tau_s \le 4.1\sp\mu$s are from pulses
    narrower than our time resolution and those that are likely to be
    free of scattering.}
  \label{scatterfig}
\end{figure}

This model was fit to the data using a least-squares minimisation and
the $1/e$ time derived from the models was taken as $\tau_s$ of an
individual giant pulse. It is known from the work of \citet{sbh+99},
that a single one-sided exponential is not sufficient to model the
complex structure of the giant pulses at this frequency. However, The
large majority of pulses in our data show that the single exponential
model agrees within $10\%$ error. Therefore, we proceeded with the
single exponential fits. The values of $\tau_s$ as a function of
observing time and their distribution are shown in the upper and lower
panels of Fig. \ref{scatterfig}, respectively. The reduction in the
scattering time towards the end of the observation is consistent with
scattered pulses tending to be dimmer, hence below the detection
threshold. Only sufficiently bright pulses are detected in the
sensitivity limited part of the observation, as discussed in
\S\ref{fcal}. The scatter tail is also not discernible from the system
noise in this part of the observation, limiting the determination of
$\tau_s$. However, there were fewer pulses so they did not contribute
to the distribution of $\tau_s$ (lower panel of Fig. \ref{scatterfig})
significantly.

The lower panel of Fig. \ref{scatterfig} shows an exponential
envelope in the distribution of $\tau_s$. The individual pulse
scattering time varies from 4.1 $\mu$s to $\sim$90$\sp\mu$s. The large
number of pulses in the distribution with $\tau_s \approx$ 4 $\mu$s is
related to our ultimate time resolution of 4.1 $\mu$s. This also
implies that a large fraction of the pulses have scattering time
$\tau_s \le 4.1\sp\mu$s. At a slightly earlier epoch than our
observations, \citet{bwk+07} determined a value of $\tau_s = 395 \pm
50\sp\mu$s at 200 MHz. Using their revised frequency scaling of
$\tau_s \propto \nu^{-3.5\pm 0.2}$, the scattering time at the centre
of our band (1373 MHz) is $0.47 \pm 0.05\sp\mu$s. At a slightly later
epoch, \citet{btk08} find a value of $\tau_s = 0.8 \pm 0.4\sp\mu$s at
1300 MHz, which contrasts with the value of 8$m$s at 111 MHz (or 1.4
$\mu$s at 1300 MHz using a $\nu^{-3.5}$ scaling law) reported by
\citet{kljs08}. With our data, we are not sensitive to scatter times
below 4.1 $\mu$s, but to the dispersion seen in the histogram of
scatter times in Fig. \ref{scatterfig} shows that variations can even
be expected within a single observation of six hours. We again refer
to Fig. \ref{GPs} for an example of the extreme form of this
variation: the different parts of the {\em{same}} pulse show different
scattering effects, imparting a significant structure to the pulse. In
their work on DISS, \citet{cr98} emphasise that considering the $1/e$
time equal to $\tau_s$ is only valid for a thin screen and does not
always hold. In light of the limited validity in interpreting the
$1/e$ time and the spread in the values of scatter times found in our
analysis, we suggest that the scattering in the direction of Crab
pulsar cannot be modelled by single thin screen. The spread in
$\tau_s$ ranges from $\le 4.1 \mu$s to $\sim 120\mu$s in our $\sim$ 6
hour-observation. This proves most of the scattering cannot be due to
the ISM, as the line of sight through the ISM does not change rapidly
enough to explain these variations. Therefore, the bulk of scattering
should orginate in the Crab nebula. The nebula can clearly give rise
to a complex screen or changes in the structures in the vicinity of
the pulsar that give rise to the short-term changes in scattering time
\citep{bwv00,lpg01,sbh+99}. The scattering of pulses cannot be in the
pulsar magnetosphere. In that case the pulses at lower frequencies
that originate higher up in the magnetosphere should show lower
scatter times, because according to the standard pulsar models, the
number density of charged particles is lower in the upper
magnetosphere \citep{lp98}. However, $\tau_s$ scales with frequency as
$\nu^{-3.5}$ \citep{pkud+06}, and this does not support the hypothesis
that scattering could have its orgins in the pulsar magnetosphere.

The diffractive scintillation timescale, $\Delta t_{DISS}$ at this
frequency was estimated by \citet{cbh+04} as 25.5s, based on pairs of
single pulses with sufficient $S/N$. However, the pulse pairs they
used were separated in time by a few pulse periods. Since our data has
good frequency resolution (224 frequency channels across 140 MHz), and
we detected several pulses with multiple components, we proceeded to
estimate possible variations in the scintillation time on shorter
timescales.

\subsection{Scintillation within single pulses}
\label{subscatter}

\begin{figure}[htbp]
  \includegraphics[scale=0.65,angle=-90]{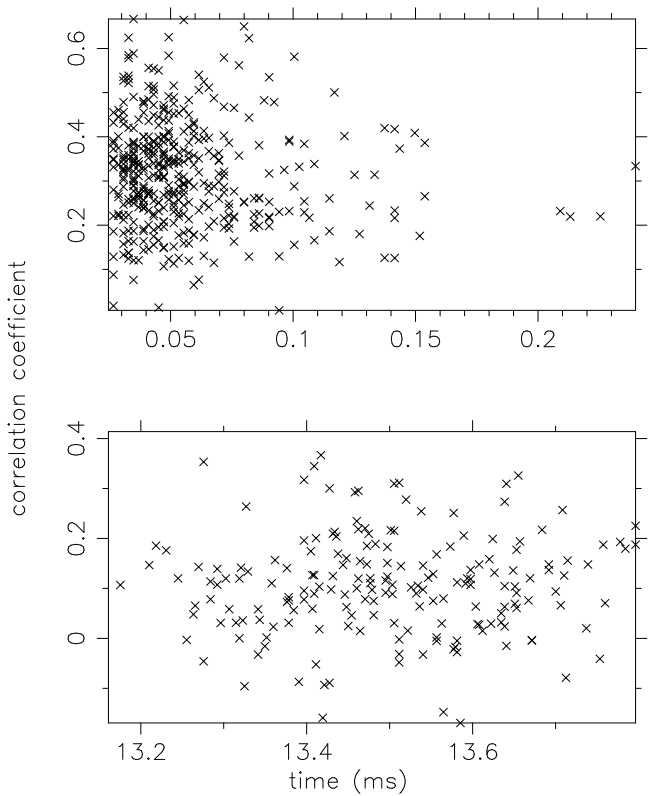}
  \caption{ Correlation coefficients of the spectra within a single pulse
  period. Top panel shows correlation between the two components of giant
  pulse, while lower panel is the double giants. The separation between the
  components $\tau$ is shown in the abscissa.}
  \label{dpeaks}
\end{figure}

The scintillation timescale within single pulses was estimated using
those pulses that show well separated components and the double pulses
discussed in \S{\ref{doublegiants}}. The search for at least two
components in single pulses was carried out based on the component
separation of $\sim$25$\sp\mu$s. This was done by examining the pulses
by eye, after an automated first pass. The first pass provided 451
giant pulse candidates, 368 of those displayed at least two distinct
shots in the main pulse phase, and 18 candidates were found in the
interpulse phase. The 197 double pulses were included in this
analysis. Assuming that the two shots of pulses are intrinsic to the
pulsar emission and that the scattering screen remains stable within a
pulse period, any scintillation would affect the two components
similarly, introducing a correlated frequency structure. The
scintillation timescale is then the $1/e$ point along the time axis of
the 2-dimensional intensity correlation function, $C(\delta\nu,\tau) =
\langle I(t,\nu).I(t+\tau,\nu + \delta\nu)\rangle$ of the spectrum
\citep{cor86}. The computed correlation coefficients between the two
components and the double pulses are displayed in
Fig. \ref{dpeaks}.

The correlation coefficient of $\sim$0.4 for many pulse component
pairs is in excellent agreement with the value derived by
\citet{cbh+04}. They derive a value of 0.33 considering the giant
pulses to be 100\% polarised, amplitude modulated, scintillated shot
noise. It also implies that these components have undergone similar
scintillation effects, ruling out the possibility of any variation in
the scattering medium on these timescales. The average correlation
coefficients computed for the double pulses is consistent with the
average value computed for the widely spaced pulse components (pulses
in the top panel of Fig. \ref{dpeaks}). Since a clear roll-off in the
values of correlation coefficient is not seen in the data presented
here, we conclude that the scintillation timescales are longer than 14
$m$s, which is entirely consistent with \citet{cbh+04}.

\section{Discussion}

To our knowledge this is the largest collection of high
time-resolution giant pulse analysis presented in the literature. Even
though some features of the giant pulse emission like the giant nano
shots are in the process of being explained \citep{he07}, several
questions still remain about the pulsar emission mechanism in general
and the giant pulse phenomena in particular. From the measured pulse
widths and the observed structure in many pulses, it is evident from
the analysis presented in this paper that the giant pulse emission is
a manifestation of temporal plasma changes in the pulsar
magnetosphere. The observed giant pulse rates are further evidence for
this temporal variation, because if the mechanism responsible for the
giant pulses is active on timescales longer than a pulse period, a
clear excess of giant pulses separated by a single rotation period can
be expected. On the basis of the giant pulse arrival times, it was
concluded that the observed giant pulse emission does not come from a
steady emission beam loosely bound to the stellar surface
\citep{lcu+95,sbh+99}. We confirm that our data do not support such a
model, for if such a beam with random wobbles operates, a
characteristic width in the giant pulses can be expected. In other
words, the distribution of the pulse widths would be normally
distributed with a mean width.

The power-law nature of the giant pulse intensity distributions was
shown by \citet{lcu+95}, and they inferred that the normal pulses
formed a separate part of the intensity distributions. In this work,
we have shown conclusively that the giant pulses consist of two
distinct populations especially for those pulses found at the inter
pulse phase.  We see a definite change in the shape of the
distribution of pulse energies as we go to lower energies and we also
see a slight broadening of the pulses. These pulses still seem to be
distinct from what might be called ``normal pulses'': they are still
narrower than most subpulses and are at least 27 times brighter than
the normal pulses. The slope of the distribution containing these
pulses is different from rest of the intensity distribution. These
pulses could possibly be the trailing part of the distribution
inferred by \citet{lcu+95}. Moreover, how these relate to the
precursor emission is unclear, which can clearly be improved upon
using the double giant pulses. While there is evidence of a broadening
of the pulses as they weaken in intensity, they do not appear to be as
broad as standard subpulses. This finding has implications in the
model derived by \citet{pet04}, where a clear power-law distribution
is explained, but not a weak giant population. The power-law index
derived also has implications for interpreting giant pulse emission on
the basis of self organised criticality \citep{btw87}, as suggested by
\citet{cai04}.

The spectral index of the Crab giant pulses reported in this work
suggests that the emission bandwidth is at least $\Delta\nu/\nu > 0.1$
and may approach the upper limit $\Delta\nu/\nu = 0.2$ predicted in
numerical models by \citet{wea98}. \citet{he07} find a similar
emission bandwidth at 9.5 GHz. Moreover, the average spectral index of
giant pulses at the interpulse phase is flatter than the giant pulses
at the main pulse phase. This possibly explains the dominant and
bright nature of interpulse giants at $\nu > 5$ GHz. We note the
prominent emergence of bimodality in the intensity distribution of the
interpulses relative to the main phase pulses. Furthermore,
\citep{he07} find upward drifting emission bands in the spectrum of
the interpulses giants and not in the main pulse giants. These
differences strongly suggest a different nature to the interpulses. To
explain the drifting emission bands, \citet{lyu07} derived an excess
plasma density of $\sim$10$^5$ and a large Lorentz factor of the
emitting particles of the order of $\sim$10$^7$, and this condition is
satisfied close to the light cylinder over the magnetic
equator. However, the model proposed by \citet{lyu07} is only valid
for $\nu > 5$ GHz, where the emission bands are observed. While
results from our observations can neither support nor rule out this
model, the difference in pulse intensity distributions we find
indicates that the interpulse giants are different in nature.

It is worth noting that the pulsar signal is a stochastic process that
contributes to the measurement noise of the pulsed intensity.  This is
especially true in the case of giant pulse emission, where pulsed flux
can exceed 1500 Jy, an order of magnitude greater than the system
equivalent flux density (SEFD) of approximately 145 Jy.
Source-intrinsic noise increases the measurement uncertainty of
various derived parameters, such as the pulsed flux density, pulse
width, scattering time, and spectral index \citet{van09}. In addition,
any temporal and/or spectral correlations -- either intrinsic to the
giant pulse emission or induced by interstellar scintillation -- will
also affect the uncertainties of any derived parameters.  The vast
majority of the pulses presented in this analysis have average flux
densities that are lower than the SEFD, and we do not expect that
self-noise will significantly alter the results of this analysis.  To
accurately quantify the impact of self-noise on parameter
distributions (such as those presented in Figures 4,5,7, and 8) would
require extensive simulations that are beyond the scope of the present
work but may provide additional insight in a future paper.

The previously unreported double pulses we found are consistent with
the occurrence rate on a purely probabilistic basis. Collecting even
more of these pulse pairs would allow for better checks of the
statistics of occurrence to ascertain that they are chance occurrences
and not indicative of some longer term underlying phenomenon driving
the giant pulse emisision. Moreover detecting more of these pulses at
higher time resolution would provide further insight into the nature
of these pulses. \citet{he07} found that the giant pulses at the
interpulse phase show an additional dispersion when compared to the
pulses at the main pulse phase. The closest pulse pair they were able
to examine were separated by 12 minutes. One may gain new insight into
the excess dispersion seen at the interpulse phase by examining the
double giant pulses, which are the closest giant pulse pair possible.

Scattering analysis of single pulses presented in this paper show a
variety of scattering times and corroborates with the analysis of
\citet {sbh+99}. They show that scattering from multiple screens or a
single thick screen is excluded because of the observed frequency
independence of the pulse component separation. From this it was
concluded that the multiple components that make up the giant pulses
are intrinisic to the emission mechanism. Using multiple components
and the double pulses, we conclude that the scintillation timescales
are greater than 14\sp$m$s, which indicates that there are no large
changes in the number density of the scattering medium along the line
of sight through the nebula on similar timescales. That the multiple
components we detect in the giant pulses are spaced by at least
25\sp$\mu$s implies that the magnetosphere and/or the plasma does not
change on these timescales, if the source intrinsic emission is less
than 25 $\mu$s. On the other hand, giant pulses may consist of
overlapping nano shots. In this case the competing models make use of
plasma turbulence leading to modulational instablity \citep{wea98} or
the induced Compton scattering of low-frequency radio waves
\citep{pet04} in the magnetosphere to explain the origin of the nano
shots. While with our data we are not sensitive to the pulses less
than 4.1 $\mu$s duration, there is an indication that the emission
bandwidth $\Delta \nu/\nu > 0.1$, suggesting that the pulses can
potentially have structure as narrow as 3.6 $n$s at this frequency.

\section{Conclusions}

The large collection of single pulses we gathered has allowed us to
perform a range of statistics with the data.  After careful flux
calibration, a detailed analysis of the pulse intensities, energies,
widths, and separation times was done by computing distributions of
these quantities. In the single-pulse intensity distributions, we find
a clear evidence of two distinct populations in the giant pulses. The
giant pulse separation times show a Poission distribution, and the
rate of occurrence of giant pulses was determined. Spectral indices
for a large number of giant pulses were computed with the narrowly
spaced multi band data. Significant dispersion in the spectral indices
was found and a small negative average spectral index was found for
the main- and interpulse giants, and they are flatter than the average
pulse emission. We also note that in some cases there is evidence for
intensity modulation with bandwidths that are smaller than the full
band but not consistent with scintillation effects. The previously
undetected double giant pulses were presented and we find that they
are not more frequent than would be expected by chance. The scatter
time for a large number of giant pulses was determined by modelling
the scatter broadening as an exponenial function and the distribution
of scatter times was computed. The double giant pulses were reported
for the first time and it is found that they are not very different
from the normal giant pulses. Using multiple emission components
either at the main- or interpulse phase and the double giant pulses, we
find no evidence of variation of the scattering material on timescales
shorter than 14 $m$s based on the correlation coefficient computed for
emission within a single-pulse period.

\begin{acknowledgements}
  The WSRT is operated by ASTRON. We thank the observers for setting
  up the observations. The PuMa--II instrument and one of us, RK, are
  funded by Nederlands Onderzoekschool Voor Astronomie (NOVA). We
  acknowledge the use of SAO/NASA Astrophysics Data System. RK thanks
  Maciej Serylak for his helpful comments. We thank the anonymous
  refree for comments that improved this paper.

\end{acknowledgements}

\bibliographystyle{aa}

\end{document}